\documentclass{webofc}
\usepackage[varg]{txfonts}   
\begin{document}
\title{AI Enabled Data Quality Monitoring with Hydra}
%
%

\author{\firstname{Thomas} \lastname{Britton}\inst{1}\fnsep\thanks{\email{tbritton@jlab.org}} \and
        \firstname{David} \lastname{Lawrence}\inst{1}\fnsep\thanks{\email{davidl@jlab.org}} \and
        \firstname{Kishansingh} \lastname{Rajput}\inst{1}\fnsep\thanks{\email{kishan@jlab.org}}
}

\institute{Thomas Jefferson National Accelerator Facility
          }

\abstract{%
 Data quality monitoring is critical to all experiments impacting the quality of any physics results.  Traditionally, this is done through an alarm system, which detects low level faults, leaving higher level monitoring to human crews. Artificial Intelligence is beginning to find its way into scientific applications, but comes with difficulties, relying on the acquisition of new skill sets, either through education or acquisition, in data science.  This paper will discuss the development and deployment of the Hydra monitoring system in production at Gluex.  It will show how "off-the-shelf" technologies can be rapidly developed, as well as discuss what sociological hurdles must be overcome to successfully deploy such a system. Early results from production running of Hydra will also be shared as well as a future outlook for development of Hydra.
}

%
\maketitle
\section{Introduction}
\label{intro}
In modern high energy and nuclear physics experiments data are costly to acquire. This makes data quality monitoring an increasingly essential component.  Failure to adequately monitor incoming data quality may result in the collection of sub-physics quality data, hindering all downstream scientific processes.  Ideally, data quality is continuously monitored and when an anomaly/fault is encountered it is immediately diagnosed and repaired with minimal experimental downtime.

\par
\label{traditional_methods}
The GlueX experiment took a more traditional approach to data quality monitoring where diagnostic plots are produced from sub-intervals of data taking known as ``Runs". Shift crews (2 individuals) are responsible for: looking over these plots regularly, responding to audible alarms, and contacting experts as needed. This responsibility is in addition to experimental configuration changes and normal data acquisition tasks. 

An alarm system is used to alert shift crews to fault conditions arising from various systems. For example, a temperature is reading above threshold.  The alarm system though has limited knowledge as to the intent of researchers and there are many fault conditions which can and do go unnoticed.  One such event that occurred during a GlueX run was a red light in a dark room being mistakenly left on after component repairs. This caused a slight increase in the background noise levels seen by the photo-multiplier tubes (PMT). It was a fairly subtle effect that showed up as an increase in the occupancy of the calorimeter blocks that were away from the beam-line. This went unnoticed by shift crews and no alarms were produced.  It is for this reason that GlueX employs additional monitoring redundancies.  These redundant measures include an Online Monitoring Coordinator (OMC) whose job it is to review, daily, plots produced during the preceding 24 hours and compile a brief as to the status of monitoring.  The entire process is labor intensive requiring many humans to operate in a regime in which mistakes are likely. \par

\section{Background}
\label{background}
\label{current_methods}
Before Hydra was developed the GlueX collaboration relied heavily on shift crews (two people in 8 hour shifts, all day, everyday, while taking data), to respond to alarms and monitor detectors, ensuring the data taken is of the highest quality.  There are many conditions for which alarm systems are not designed, or cannot be easily designed to handle.  Such conditions require human oversight and recognition.  To aid in this type of monitoring GlueX developed and deployed RootSpy, a system which monitors the data stream, gathering and displaying statistics from the various detector systems primarily in the form of histograms.  Shift crews are tasked with regularly monitoring a rotating set of histograms produced from the accumulating statistics of each run.  Conditions, such as electronics cutting out during data taking, may not be detectable through these plots.  This is especially true for intermittent problems which do not significantly effect the plots of integrated statistics.  To correctly detect such problems RootSpy was developed to allow a user to ``Reset" a histogram.  This process involves clearing a histogram and regenerating it with newly gathered statistics. After ensuring that no short timescale problems existed the histogram could be ``Restored".  Restored histograms would then contain totality of all collected statistics accumulated for the current run.  There were no controls in place to ensure that shift crews actually analyze any of the plots produced by RootSpy.  Nor was there any guarantee that plots would be reset and inspected;  informal polling indicated that with the other burdens on shift crews this type of monitoring was not commonly performed. \par
\label{rootspy_img}
Each plot which is integrated over the entire run and produced by RootSpy is archived in Portable Graphics Format (PNG).  These images could then be displayed on the web via webpages designed specifically for collation of these images.  Daily emails are automatically dispatched with the expectation that system experts would then review these images and take necessary actions to correct any issues discovered.  This amounts to close to 1,000 individual plots a day which must be scrutinized. The OMC managed the monitoring webpages, emails, and was responsible for reviewing all plots to produce a daily brief on the status of the data taking.  Given the volume this was rarely, performed in its entirety.  The OMC typically relied on a few core plots.
\section{The Hydra System}
\label{hydra}
Humans are involved with every step of data quality monitoring and a lack of methods for ensuring that every step is properly taken leads to a system fraught with the possibility that issues go unrecognized.  Therefore a need exists to introduce methods for monitoring that are automated and persistent.  With this in mind Hydra was developed.\par
\label{what_hydra}
Hydra is an Artificial Intelligence (AI) system, based on Tensorflow, designed to supplant humans in monitoring plots produced by RootSpy. The decision was made early on to utilize the developments in computer vision as well as ``off-the-shelf" components to analyze the PNG files directly.  The images were chosen, over the underlying histograms or raw data, for several reasons.  Firstly, development time was drastically reduced as all of the previous images were already stored, collated, and readily accessible; in stark contrast with using the underlying data/histograms, which would be very labor intensive and require the creation and maintenance of custom, GlueX specific, code.  Secondly, humans perform these tasks by looking at images.  Utilizing similar methods to humans should make the system more interpretable, allowing for easier diagnosis of Hydra in the future.  Thirdly, using images directly makes Hydra more generic; each data instance may be derived from a specific format, but all can be transformed into an image (e.g. saving a histogram as an image in Root). Finally, by utilizing the PNG image files directly we can leverage all of the advancements in computer vision, image classification, and semantic segmentation.

\subsection{Inception v3}
\label{Inception}
The Inception v3 network\cite{szegedy2015rethinking} is a very powerful deep learning model developed by Google for classifying images. It is the third iteration of the Inception series having more than 78\% accuracy on the image-net data set\cite{deng2009imagenet}. It is  comprised of a very complex interconnection of convolutional layers, pooling layers, dense layers, and custom inception layers forming a deep neural network with over 50 layers. Inception v3 also utilizes factorized convolution and label smoothing\cite{szegedy2015rethinking}. Since it is included in Keras\cite{chollet2015keras} and Tensorflow\cite{tensorflow2015-whitepaper} packages in python, someone with little knowledge in AI/Machine Learning can take it off the shelf and train with their own image data sets for classification.  We chose inception V3 because of it's assurance of good performance on image classification proven by its performance on the image-net data set.  This allowed us to focus on building the Hydra system instead of having to worry about model design and development.  Because Inception v3 is a deep and complex network it is expected to require a large set of labeled data.

\subsection{Data Labeling}
\label{labeling}
Deep Learning algorithms, such as Inception v3, typically require a large number of labeled data.  Gathering training data is not problematic in the case of Hydra because all necessary data had already been gathered.  Labeling said data, however, poses a challenge. There exists approximately 30,000 examples of each plot type analyzable by Hydra.  In total the initial roll-out of Hydra contained approximately 250,000 images which required labeling by detector experts.  Detector experts' time is valuable and convincing anyone to label 30,000 images is no easy feat.  Therefore, a system for efficiently labeling sample images was needed.\par
\label{hydra_labeler}
For the purpose of labeling we developed a web page, dubbed ``HydraLabeler".  This web page is accessible from anywhere in the world, perfect for the detector expert on the go. It does, however, require the proper credentials to log in.  Permissions are granted to individuals to perform labeling on a plot-type basis.  This ensures there is no accidental mislabeling of plots by people not deemed an expert for that particular plot type.\par
\begin{figure}
\centering

\includegraphics[width=14cm,clip]{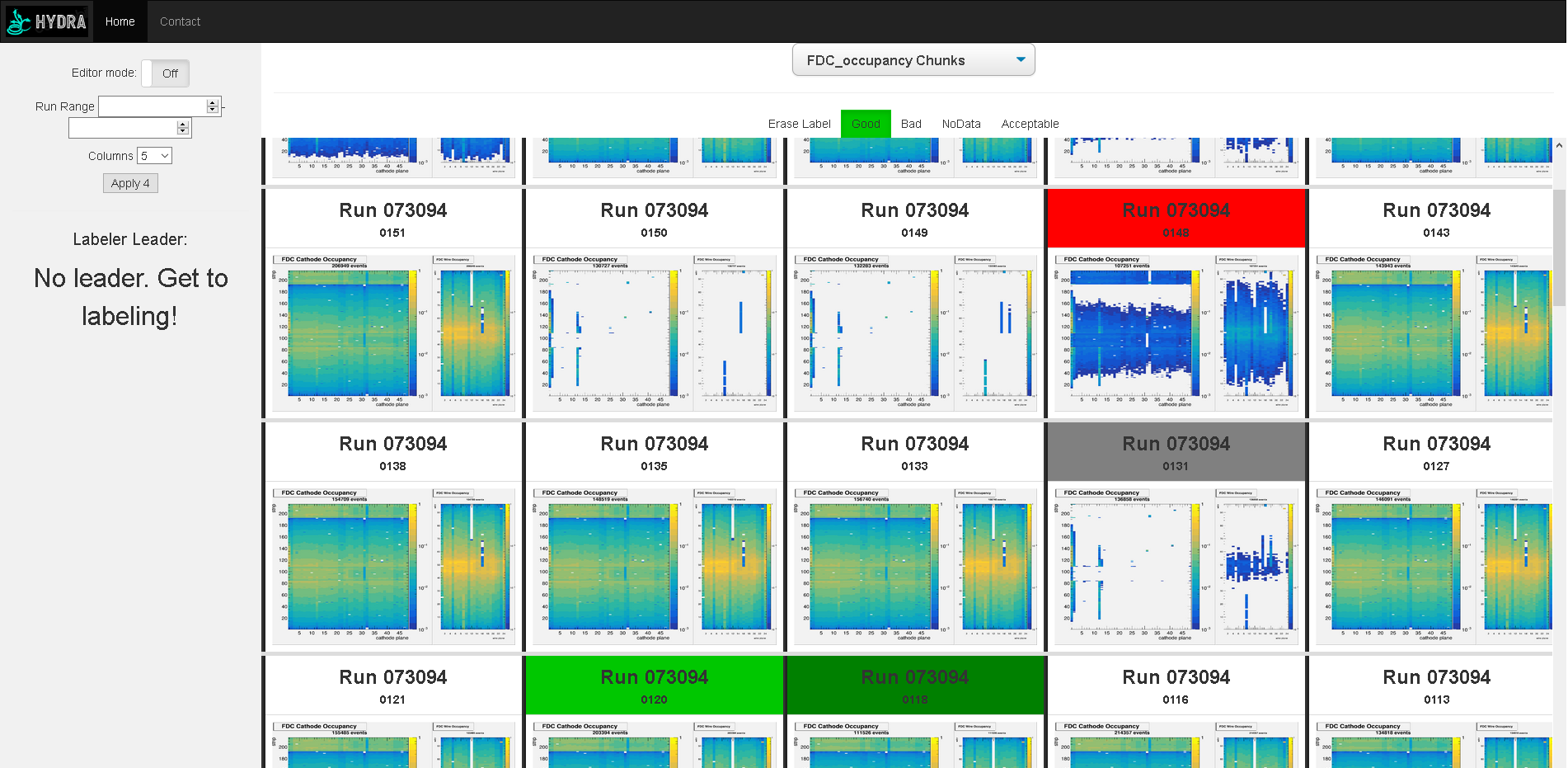}
\caption{A screen shot of the HydraLabeler web page for an arbitrarily chosen plot type. It shows a grid of unlabeled images with some candidate labels, from the label palette at the top, waiting to be applied.}
\label{labeler}   
\end{figure}

\label{labeler_gui}
Once logged in and plot type selected the user is presented with an array of unlabeled plots, chronologically ordered row by row, as well as a palette of labels each with a unique color (See Figure \ref{labeler}).  A candidate label is chosen from the palette at the top of the page. The user then ``paints", by left-clicking, desired images with the chosen label. By holding down the ``Shift" key, after labeling an image, and labeling another image, a user can label all images in between with the chosen label.\par
\label{barrier_reduction}
This style, coupled with the fact that detectors are designed to stably operate for extended periods of time allows for the labeling of hundreds of images in a matter on a couple of minutes.  This drastically reduces the barrier to labeling images for training and saves experts' time as well as ensures that Inception v3 has a sufficiently large labeled set for training.\par

\subsection{The Hydra Database}
\label{hydra_labeler}
References to all of the images, as well as all of the labels submitted via the web page, are stored in a database.  This database is divided into two arms.  The first is a complete record of all logged image files gathered from the existing RootSpy system.  It provides the basis for the training and testing of any models included in the Hydra system. The second arm of the database supports the operational functions of Hydra, these functions include logging and monitoring of the Hydra system.

\subsection{Training}
\label{hydra_training}
References to each image file are stored by recording all of the necessary meta-data to reconstruct the full path to the file itself.  These include the file system the files are on, the run period the image belongs to, the run number of the run the image belongs to, and the file name itself.  This was done instead of merely storing the full paths to allow for easier correlation with other data sources (e.g. it is known that run 12345 had a detector in a bad state). \par
\label{hydra_dataframe}
During training, data is retrieved from the database and full paths reconstructed and ultimately stored in a pandas dataframe. Each image can be weighted so that it appears many times in the training dataframe.  This is important if some rare case has too few examples to pull the model weights sufficiently to learn.  This dataframe is then split into training and validation sets and shuffled.  For the basic categorizations the variance in ``Good" plots and plots providing ``No Data" (when the RootSpy system was down) were much much less than the variance in ``Bad" images, where a whole host of problems may manifest themselves with wildly different symptoms.  In order to not have plots indicating problems washed out Hydra employs ``strategic under-sampling" in which images from statistically larger categories are randomly thrown out in favor of the smaller categories with higher variance.  This is naively counter intuitive but provided models which were as accurate or more accurate in far less time.  Generally, the train-validation split heavily favors training data, with upwards of 95\% of data being reserved for training.  Unlike training with most image classification algorithms, where it is highly likely the model will be asked to make inferences on unique images, Hydra will be processing very standard images.  The images are comprised of the same plots, with standardized text in the same places, with very little variance in overall image properties.  As there is no expectation of high variance in images during inference erring on overfitting is preferred.  Each model takes approximately 4 hours to train with strategic under-sampling and is between 93 and 99 percent accurate (depending on plot type).\par

\subsection{Testing}
\label{model_testing}
After a model has been trained it is recorded in the database and inference over all valid plot types is performed.  Each confidence in each category for each image is also recorded in the database.  A report is generated showing the images to which the model and ground truth labels are in disagreement.  This information is then passed along to detector experts for review.  Often human errors in labeling are discovered and corrected at this stage.  If enough image labels are updated then a new training would be performed.  If the AI was right and the human wrong in a majority of cases the labels are simply updated for future training/analysis. The storage of the confidence in all the labels for every image allows for a detailed analysis to be performed before a model is deployed.  Because each image's label is also recorded a confidence level confusion matrix is trivially created.  This augmented confusion matrix contains not only the counts in each matrix element but also shows the distribution of the confidence the system had in making its determination.  This allows one to answer higher order questions such as, ``When the model is wrong is it confident in its erroneous determination or not?".  Additionally, excesses in confidence distribution could point to potential bifurcation in categories or fundamentally different modes of detector operation.  With these distributions thresholds can be established to control the true rate of false positive alarms by dialing in the sensitivity.  This freedom addresses one of the concerns about the deployment of an AI system, discussed later\ref{hydra_operation}.

\subsection{Monitoring}
\label{hydra_monitoring}
Hydra provides feedback as to the state of detectors in two ways, a web-accessible real-time view of the status of the detectors, and a trailing 24 hour view of every plot Hydra deemed to be bad or was under confident in.  The real-time view is based on a security camera type layout in which the last image to be inferred of each plot type is displayed along with its classification and confidence, all in one window.  If a ``Bad" state is detected the image frame is highlighted and text color changed on the web page, notifying shift crews to pay attention to the detector system or take action to resolve the issue.  The trailing 24 hour view allows overall monitoring of the detector status.  This is meant to allow detector experts to more easily review the prior day of running.  Instead of looking through all plots experts need only concern themselves with a subset where there may be some concerns.  Because Hydra analyzes in finer time windows than humans (hydra checks plots on the order of once every minute) it can easily detect intermittent issues that get washed out in a run's total statistics.  This can inform future operations, for example, a piece of electronics maybe in the process of failure by intermittently turning off an on.  In this case, GlueX could take advantage of an access to the experimental hall to replace the component before a complete failure occurs.  This would prevent a future interruption to data taking, allowing GlueX to collect more statistics.

\section{Operation}
\label{hydra_operation}
The Hydra system receives images from the current RootSpy system and matches them up to the appropriate preloaded model.  It then logs its actions to a local file and records the image processed as well as the classification and confidence in that classification.  This information is stored in the database and provides shift crews with actionable information.\par
\label{concerns}
One concern collaborators had in the operation of Hydra was the overall alarm rate.  False positives could drive shift crews to turn off or otherwise ignore Hydra.  In this case any benefits provided by Hydra would be rendered moot.  To combat this concern confirmation thresholds are set for every class of every model independently. This allows each classification to only trigger an alarm when Hydra is confident enough in a classification that is supposed to trigger an alarm.  In cases where the confidence falls below threshold no alarm is raised, the image is, however, saved and flagged for future labeling and training.\par
\label{assurances}
Another concern was one of adaptability,  the collaboration needs some assurances that the system would not be biased and changes in nominal running conditions not lead to missed problems or false alarms.  While running Hydra collects a tune-able unbiased sample (a fixed percentage of all images gathered). This sample can be labeled and included in future training or spot checked for accuracy, ensuring that Hydra's performance is always scrutinized.  This spot checking adds very little burden to collaborators as only a handful of plots a day need be analyzed instead of the hundreds of images per day before.\par
\label{earning_trust}
Ultimately, Hydra must earn its trust and prove its utility by spotting problems missed by shift crews in situ. On 08/27/2020 Hydra did exactly that.  During a run half of the high voltages in one detector were off as were all of the positive high voltages of another.  No alarm was raised for either issue, likely due to a communication issue with the alarm handler itself. In the case where half of the high voltages were off the occupancy plot was visibly bad; in the case where only the positive high voltages were off changes to the occupancy plot were subtle enough not to be detectable by the shift crew.  Hydra, however, managed to spot both problems immediately and began flagging the problematic detectors (Note: audible alarms had not been set up for Hydra yet due to the concerns about potential alarm rates).  One expert even noted that Hydra alerted on an issue that they themselves would not have spotted by eye. This incident showed the power of an AI system and led to a new rquirement that Hydra be up and running whenever data is being taken.

\section{Future Developments}
\label{future}
Our vision with Hydra is to make it fully capable of pin pointing not only problems but diagnosing the exact issues in the detector and/or data collection system.  We plan to elaborate on the "Bad" category, splitting that label into different categories or subcategories such as "Bad:board-1 out", "mismatch between running conditions and data" etc.  With this specificity Hydra will then be able to prescribe or even perform corrective actions on its own (e.g. power cycling a component).\par
Hydra will also be optimized. So far Hydra has taken the same model, Inception v3, and the same training parameters to train all of its models.  A suite of model architectures will be made available, chosen and optimized for the data being processed.  These optimizations will decrease the total time it takes to train a model, decrease inference time by reducing model depth and complexity, and produce more generalized models.\par
To facilitate the application of Hydra to other use cases Hydra will be packaged as a complete system. Potential users will be able to install Hydra with minimal configuration and have access to an API to develop experimental specific parts of the system.  These parts include, the handling of the data and the actions that Hydra should take given specific inferences.  Work is already underway creating python classes which can be and have been utilized by the GlueX specific implementation of Hydra.


\section{Summary}
The Hydra system employees off-the-shelf AI technologies to implement a data monitoring system that supplements one of the most tedious tasks currently performed by shift crews and Online Monitoring Coordinators. Utilizing Google's Inception v3 network for image classification, the system is able to view plots at a much higher frequency and classify them with much higher consistency than human shift crews. Early anecdotal evidence demonstrated its ability to catch issues during live data taking that humans missed. A key aspect of the success of the system is a user friendly interface developed so that system experts can label large numbers of plots very quickly that are then used to train models to replicate the expert's judgement. The system was developed within the GlueX experiment but is designed for use with any experiment. 
 
\bibliography{bibliography}
\end{document}